# Hierarchical fringe tracking

Romain G. Petrov[*a], Thami Elhalkouj[b], Abdelkarim Boskri[b], Jean-Pierre Folcher[a], Stephane Lagarde[a], Yves Bresson[a], Zouhair Benkhaldoum[b], Mohamed Lazrek[b], Suvendu Rakshit[a].
[a]Laboratoire Lagrange, UMR7293, Université de Nice Sophia-Antipolis, CNRS, Observatoire de la Côte d'Azur, Bd. de l'Observatoire, 06304 Nice, France;
[b]Université Cadi-Ayyad, Faculté des Sciences, Laboratoire de Physique des Hautes Energies et d'Astrophysique, 40000, Marrakech, Maroc.

## ABSTRACT

The limiting magnitude is a key issue for optical interferometry. Pairwise fringe trackers based on the integrated optics concepts used for example in GRAVITY seem limited to about K=10.5 with the 8m Unit Telescopes of the VLTI, and there is a general "common sense" statement that the efficiency of fringe tracking, and hence the sensitivity of optical interferometry, must decrease as the number of apertures increases, at least in the near infrared where we are still limited by detector readout noise. Here we present a Hierarchical Fringe Tracking (HFT) concept with sensitivity at least equal to this of a two apertures fringe trackers. HFT is based of the combination of the apertures in pairs, then in pairs of pairs then in pairs of groups… The key HFT module is a device that behaves like a spatial filter for two telescopes (2TSF) and transmits all or most of the flux of a cophased pair in a single mode beam. We give an example of such an achromatic 2TSF, based on very broadband dispersed fringes analyzed by grids, and show that it allows piston measures from very broadband fringes with only 3 to 5 pixels per fringe tracker. We show the results of numerical simulation indicating that our device is a good achromatic spatial filter and allowing a first evaluation of its coupling efficiency, which is similar to this of a single mode fiber on a single aperture. Our very preliminary results indicate that HFT has a good chance to be a serious candidate for the most sensitive fringe tracking with the VLTI and also interferometers with much larger number of apertures. On the VLTI the first rough estimate of the magnitude gain with regard to the GRAVITY internal FT is between 2.5 and 3.5 magnitudes in K, with a decisive impact on the VLTI science program for AGNs, Young stars and planet forming disks.

## 1. INTRODUCTION

### 1.1. The limits of Optical Long Baseline Interferometry

Optical Long Baseline Interferometry (OLBIN) is in a decade on intensive astrophysical production, and the next 5 to 10 years are promising even better results with the VLTI 2nd generation instruments GRAVITY (K band) and MATISSE (L, M and N bands) and the continuing development of the US interferometers CHARA and MPOI in the visible and the near infrared. However the range and interest of astrophysical applications of OLBIN are limited by:
- The limiting sensitivity, as the limiting magnitude is currently around K~10 with the UTs and the best plans already accepted by ESO do no not foresee going beyond K~10.5 with GRAVITY. The limit on smaller apertures such as the CHARA 1m telescopes or the VLTI 1.8m ATs ranges between 6 and 8 in the K band. This is a strong limit for at least two key science programs:
  o The observations of AGNs and QSOs, in particular at the spectral resolutions needed to resolve the BLRs[1];
  o The imaging of star and planet forming disks with relocatable small telescopes.
- The accuracy of interferometric measures:
  o The accuracy of absolute visibility is a key parameter to constrain the overall angular size of the source, particularly when the source is fairly unresolved. It is also critical to detect low contrast features on fairly resolved sources.

---

[*] romain.petrov@unice.fr; phone : +33 4 92 00 39 61; lagrange.oca.eu

- o The accuracy of differential phase is crucial to observe very unresolved sources. It is also a key parameter to image low contrast features visible in specific spectral channels.
  - o The accuracy of closure phase is critical for the reconstruction of complex images with low contrast features, such as subtle planet forming tracks in circumstellar disks.
- The imaging capacity, which is currently mainly limited by the u-v coverage, i.e. the number of apertures and the possibility to relocate them.
- The angular resolution, needed to observe smaller sources or smaller features, which can be improved by increasing the baselines or reducing the observing wavelength.

There is a debate inside the OLBIN community about the order of priority about progress on these limits. In fact, they are strongly interconnected. Improving the sensitivity permits looking at smaller targets and hence needs either higher resolution or more accurate measurements. The number of apertures improves the u-v coverage, the imaging capacity and the number of simultaneous measures, which can partially compensate the accuracy on individual measures, but getting the highest dynamics of a given u-v coverage and for a given image complexity requires progress in measurement accuracy. The common sense is that there is a necessary trade off between number of apertures and sensitivity. We describe in this paper a fringe tracker architecture that solves this contradiction.

Discussing all the limits of OLBIN and the best tradeoff between all possible improvements depends on the specificities of each astrophysical program, and a global analysis is well beyond the scope of a single SPIE paper. Here we will concentrate on possible improvements of the fringe trackers and of the limiting sensitivity.

**1.2. Limiting magnitude in Optical Long Baseline Interferometry**

The limiting magnitude of an OLBIN instrument can be set by different parameters, related to the detection and the stabilization of the fringes, or to the accuracy on given interferometric observables at some spatial frequencies $\lambda/B$, or the accessible resolution elements (resels) and dynamics of a reconstructed image.

Here we focus on the capacity to detect and stabilize the fringes. If the fringes are stabilized within a fraction of wavelength $\lambda$, we say that the instrument is cophased. Then it is possible to integrate for a time much longer than the fringe coherence time due to the atmosphere. When we can only maintain the interferometer OPD inside the coherence length $C = \lambda^2/\delta\lambda = \lambda R$, we say that the interferometer is coherenced. If the coherence length $C$ is large enough to make sure that the fringes are present during all the time needed for a science source observation between that of two brighter calibrators, i.e. typically for 15' to 30', than we can observe in the so-called "blind mode". This blind mode is accessible at large wavelengths $\lambda$ or large spectral resolutions $R$. In coherenced or blind mode, the exposure time of individual frames must be shorter than the fringe coherence time to avoid fringe shifts (the so called piston jitter) to destroy the fringe contrast, or to decrease the "instrument + atmosphere" contrast stability and hence the absolute visibility accuracy.

Fringe detection and stabilization can be achieved by different instruments and on different targets.
- We can use the science instrument itself on the science source. This is the way AMBER[2] and MATISSE[3] work in low spectral resolution. The instrument can be a good coherencer, like AMBER in the AMBER+ mode, but the trade-offs imposed by its science specification, such as, for example, the necessity to read a large number of pixels for a full analysis of all spectro-interferometric measures, very often prevents its optimization for faint sources observations.
- The fringes can be stabilized on the science target, by a fringe tracker (FT) collecting a fraction of the source flux, if possible in a spectral band different from the science one. So far, the FTs in the infrared are all cophasers while in the visible they are coherencers. High accuracy cophasing implies extremely short exposure times and this is the main reason for the severe FT limiting magnitudes.
- The FT or the science instrument can be cophased or coherenced on an off-axis reference source. This reference source must be close enough for the atmospheric piston difference (the anisopistonic error) introduced by the angle between the source and the reference to be smaller than the fringe stability requirement. The probability to find a reference source brighter than the FT limiting magnitude in the isopistonic angle defines the sky coverage. With the current limiting magnitudes the sky coverage is so poor that off-axis fringe tracking is limited to a handful of targets or specific topics such as the study of the galactic center with GRAVITY[5].

The OLBIN common sense in the near and thermal infrared is that:

- Science instruments can do the fringe detection in low spectral resolution ($R \ll 50$), but even for this mode the stability of the instrumental contrast and hence the accuracy of absolute visibility is very substantially improved by the use of a fringe tracker.
- The sensitivity of observations at any higher spectral resolution ($R > 50$) is set by this of a FT allowing to stabilize the fringes and to achieve frame exposure times long enough for the photon noise to get higher than the detector noise in spite of the high spectral dispersion of the source flux. For example, in current GRAVITY plans, all higher spectral resolution modes will be possible only below the limiting magnitude of its internal FT, which is of the order of K~10.5.

We have shown recently, theoretically and experimentally that the limiting magnitude for medium resolution (MR) observations can be higher than this of a FT with special observation modes (~blind observations) and the corresponding specific coherencing algorithms[4], particularly if the instrument is optimized for these MR observations like OASIS[1]. However, even for these instruments, the measurement accuracy is much better if the fringes can be cophased. The Rakshit & Petrov's paper[1] in these proceedings illustrates the gain introduced by such a FT for the observations of AGNs with all VLTI instruments, including GRAVITY, if it can reach a limiting magnitude K>13.

**1.3. The parameters of the Fringe Tracking problem**

A fringe tracker must stabilize the fringes within $\lambda/n$, with $n > 20$ for good absolute visibility accuracy, and $5 < n < 10$ if we want to boost the sensitivity and are happy with the accuracy of color differential measures. The first condition for this stabilization is that the fringe sensing error on the piston $\sigma_p$, and the corresponding phase $\sigma_\varphi$ error are:

$$\sigma_p = \frac{\sigma_\varphi \lambda}{2\pi} \ll \frac{\lambda}{n} \text{ thus } \sigma_\varphi \ll \frac{2\pi}{n} \quad (1)$$

For "all in one" multi-axial fringe sensors, like in AMBER and MATISSE, the phase error $\sigma_\varphi^m$ is given by:

$$\sigma_\varphi^m = \frac{\sqrt{n_T n_* \tau + n_p \sigma_R^2 + n_T n_{th} \tau}}{n_* \tau V_I} \quad (2)$$

- $n_T$ is the number of apertures;
- $n_*$ is the number of source photons from each aperture contributing to the interferogram; here we assume, for simplicity, that all apertures contribute with the same number of photons;
- $\tau$ is the exposure time of any individual measurement used by the fringe sensor;
- $n_p$ is the number of pixels, or of measures, needed for a piston estimate; in multi-axial instruments $n_p$ is typically given by $n_p = a n_\lambda \frac{n_T(n_T-1)}{2}$ with
   - $2 < a \leq 4$ is the number of measures per fringe peak and spectral channel;
   - $\frac{n_T(n_T-1)}{2}$ is the number of baselines;
   - $n_\lambda$ is the number of spectral channels necessary for an unambiguous determination of the piston; the minimum value is $n_\lambda=3$ but $n_\lambda = 5$ is a quite typical value;
- $\sigma_R$ is the rms detector read-out noise per pixel or per measure;
- $n_{th}$ is the thermal background contribution, per beam;
- $V_I$ is the instrumental contrast.

For a pairwise fringe sensor, similar in design to the PIONIER[6] instrument or to the internal GRAVITY fringe sensor, the flux from each telescope is divided between $n_T - 1$ pairs of telescopes. Then the phase error is:

$$\sigma_\varphi^p = \frac{\sqrt{\frac{2(n_* + n_{th})\tau}{n_T - 1} + n_p' \sigma_R^2}}{\frac{(n_* + n_{th})\tau}{n_T - 1} V_I} \quad (3)$$

with $n_p' = a n_\lambda$ where $n_\lambda$ is the number of spectral channels, as in the multi-axial case, or the number of broadband fringes that must be scanned to get an unambiguous piston.

- The thermal background $n_{th}$ can be managed in different ways, for example if the beam separations or combinations are made inside or outside the cryostat. However, as fringe sensing is usually made with broadband near infrared measurements, we will neglect this term in the following, because $n_{th}\tau \ll n_p \sigma_R^2$.
- The readout noise is a characteristic of the detector. We will assume that all the fringe sensors we discuss use the same near infrared detector with the best current value $\sigma_R \simeq 3e^-$.
- The instrumental contrast can be optimized by the FS design, but we will assume that all systems have the same high $V_I \simeq 0.9$.

- The flux collected from each telescope $n_*$ is strongly affected by the interferometer and instrument design, as they change the overall transmission. For this paper we will assume that this transmission is globally the same for all designs, except for some specific filter transmission $T_f$ and for the global bandpass $\Delta\lambda$ used in each spectral channel, which can be very different for integrated optics and bulk optics: $n_* \propto n_{g*}\Delta\lambda T_f$. $n_{g*}$ is a function of the source magnitude.
- The exposure time $\tau$ is a very critical parameter, set by the atmospheric piston coherence time (10 to 100 ms in the infrared), by the frequency of the telescopes and interferometer perturbations and vibrations and by the robustness of the control loop. The latest is a major cause for sensitivity limit, as control loop robustness in the presence of instrument vibrations imposes $\tau$ (much) smaller than 1 ms, well below atmospheric piston coherence time. Another paper in these proceedings[7] discusses control loop work intended to optimize $\tau$, but we will assume here that this applies equally to all concepts, and that we can use the same $\tau$ for all.

Usually, at the fringe-tracking limit, we are dominated by detector noise and we have $n_T(n_* + n_{th})\tau \ll n_p\sigma_R^2$ and $\frac{2(n_*+n_{th})\tau}{n_T-1} \ll n'_p\sigma_R^2$. Then, the phase errors $\sigma_\varphi^m$ for the multi-axial and $\sigma_\varphi^p$ for the pairwise FT are:

$$\sigma_\varphi^m = \frac{\sqrt{\frac{n_T(n_T-1)}{2}}\sqrt{an_\lambda\sigma_R^2}}{n_*\tau V_I} \text{ and } \sigma_\varphi^p = \frac{(n_T-1)\sqrt{an_\lambda\sigma_R^2}}{n_*\tau V_I} \quad (4)$$

This gives a small advantage $\sqrt{\frac{n_T}{2(n_T-1)}}$ to the multi-axial scheme, but the key point is that in both cases the performances decrease with the number of telescopes. Even if we consider that we have $n_T(n_T-1)/2$ measures to compute $n_T-1$ pistons, we confirm that in the two classical approaches there is a conflict between the imaging capability, i.e. the number of apertures, and the sensitivity.

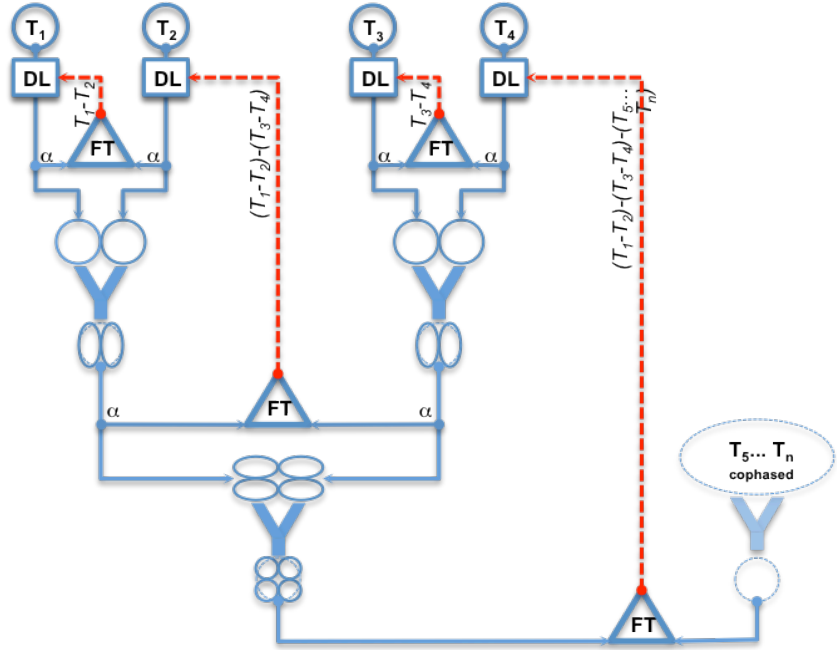

Figure 1: The concept of Hierarchical Fringe Tracking. We cophase pairs of telescopes, then pairs of pairs, pairs of groups, etc. Each fringe tracker FT sees two incoming beams, and leaves a fraction α>50% of the flux of each beam to cophase lower level groups of telescopes. Each FT drives only one delay line (DL). There are solutions for any number of apertures but the best optimization can be achieved for $2^k$ telescopes. Then, all FTs receive $2^k f_T/(2^k - 1)$ photons.

## 2. CONCEPT OF HIERARCHICAL FRINGE TRACKING

Hierarchical Fringe Tracking[8] is intended to overcome the decrease in sensitivity with the number of apertures. The initial concept is illustrated in figure 1. We first cophase pairs of telescopes, then pairs of pairs, then pairs of groups of apertures, in $int[lg_2(n_T)]$ levels. Each fringe tracker sees two incoming beams, which can come from a single aperture or from any number of cophased apertures, and drives a single delay line.

*If, at each FT level, a fraction α=0.5 of the flux is transmitted for cophasing the lower levels, each fringe tracker sees the number of photons $f_T$ produced by a single aperture, and the performance is independent of the number of apertures.* However, as the lowest level FT can use all the flux, it is possible to have α>0.5. For $2^k$ telescopes, simple algebra shows that all FT are fed by the same flux $f_T 2^k/{2^k-1}$ if $\alpha = 2^{k-1}/{2^k-1}$. Then a 2T fringe tracker is indeed more efficient, as it sees $2f_T$ photons, but as $k$ increases, the flux feeding each FT very rapidly converges toward $f_T$ and never goes beyond. As it is possible to cophase a pair of apertures with a single telescope and to cophase groups containing different numbers of apertures, the concept is not limited to arrays with $2^k$ telescopes, even if such arrays are easier to optimize.

A key issue is to merge the beams left for lower level FT in a way that makes them identical to a single aperture beam. This can be achieved with Y couplers in integrated optics, which will be efficient if the merged beams are in phase. This would limit the spectral band or impose to use a set of single band integrated optics systems. There are also bulk optics solutions, with an example discussed in the next section, but merging beams resulting from groups with different compositions might introduce pupil shape differences and then coupling efficiency issues.

### 3. A SPATIAL FILTER FOR TWO TELESCOPES

Spatial filtering is a key issue in optical interferometry, as it improves very substantially the instrumental contrast and stability. It is also central to fringe tracking, as the FT must measure the global "piston" i.e. difference between the mean OPDs in each beam, whatever the other OPD perturbations are. Spatial filtering is usually performed using single mode fibers, like in AMBER, or integrated optics like in PIONIER and GRAVITY, but the study of MATISSE has demonstrated that a succession of image plane "pin holes" and pupil plane masks can have good spatial filtering properties with a good coupling efficiency[9].

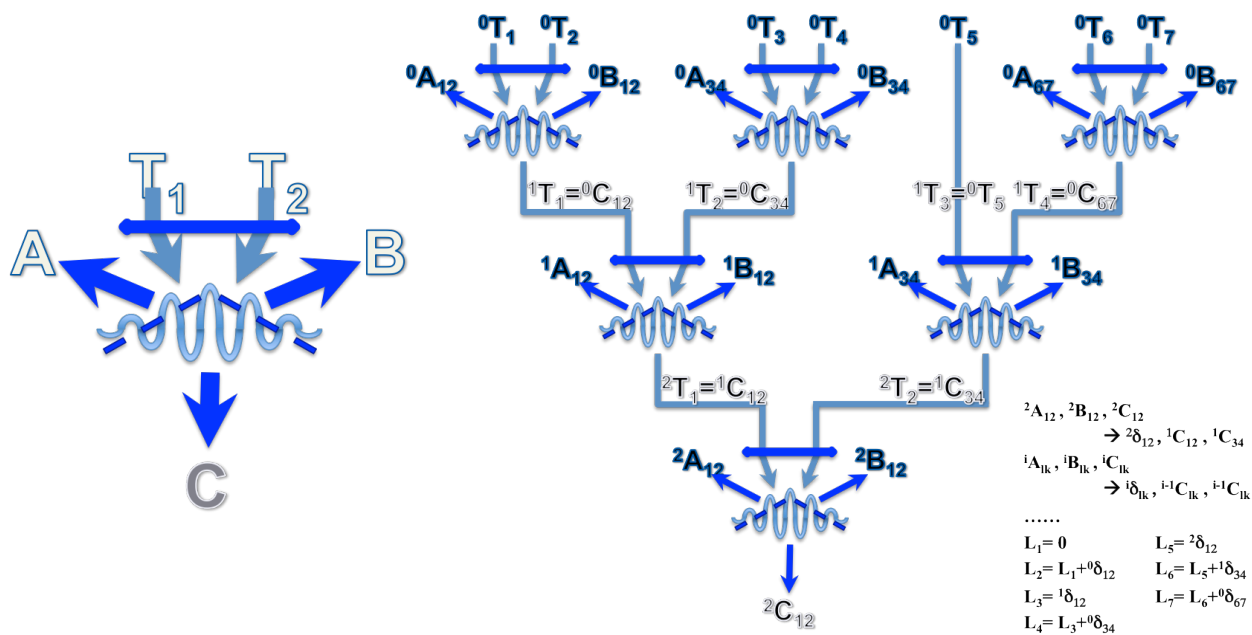

Figure 2: Functions of a two telescopes spatial filter (2a: left) and their combination in a hierarchical fringe tracker (2b: right). The two beams coming from $T_1$ and $T_2$ are transmitted as a single mode beam in C, when $T_1$ and $T_2$ are cophased, with an efficiency *β>0.5* (ideally *β=1*). When the incoming beams are not cophased, the flux is reflected in beams A, B… that are used to measure the piston difference between beams $T_1$ and $T_2$. This piston measurement uses *all* the flux produced by $T_1$ and $T_2$ as it uses fully A, B and C. When several filter are combined in a hierarchical level (figure 2b, left), at the lowest level, A, B, and C are used to compute the piston between the two beams and the flux in each beam, which are the "C" for the above level, which allows propagating this computation to all pistons. The level $k$ FT see a flux $f_k = 2^k \beta^{k-1} f_T$ that increases with $k$ as soon as *β>0.5*.

The concept of Hierarchical Fringe Tracking can be improved if we can design a Two Telescopes Spatial Filter that:
- o Transmits all (or most of) the flux when the two incoming beams are cophased, in a single mode beam, i.e. with a single global OPD.
- o Reflects or deflects the flux toward detectors that can be used to measure the differential piston when the incoming beams are not cophased.

The concept of such a spatial filter is illustrated in figure 2a, and their combination in a hierarchical fringe tracker is shown in figure 2b. The two beams coming from $T_1$ and $T_2$ are transmitted as a single mode beam in C, when $T_1$ and $T_2$ are cophased, with an efficiency $\beta>0.5$ (ideally $\beta=1$). When the incoming beams are not cophased, the flux is reflected in beams A, B… that are used to measure the piston difference between beams $T_1$ and $T_2$. A piston estimator combining A, B and C, uses *all* the flux arriving at that level.

When several filter are combined in a hierarchical structure (figure 2b), at the lowest level A, B, and C are used to compute the piston between the two beams and the flux in each beam, which are the "C" for the above level, which allows propagating this computation to all pistons.

A fringe tracker at the level $k$ sees a flux $f_k = 2^k \beta^{k-1} f_T$ that increases with $k$ as soon as $\beta>0.5$. The most critical FT are these of the first level that use a flux $f_1=2f_T$.

The bootstrapping capability of this approach has not been fully analyzed and it is easy to imagine that it depends both from the source and the array configuration. However, this concept seems quite favorable because:
- The telescopes can be grouped in order to avoid the longest baselines, as the piston of each group can be set to this of any of the individual apertures, thus allowing choosing the closest apertures between the two groups.
- The pairs of the first level, that have the lowest fundamental (photons versus detector noise) SNR, can be built by relatively close apertures, while the longest baselines will be associated with the higher SNR lowest levels.

The control loop of such a system is being investigated and it is too early to discuss its potential and problems. However, as all pistons are computed exactly from one set of measures, there is no reason to have time requirements more severe than in the other fringe trackers. We even have the flexibility to be stricter on individual exposure times for the lowest levels with the longest baselines and hence possible highest piston jitter but also the highest fluxes.

## 4. A DISPERSED FRINGES BROADBAND ACHROMATIC 2 TELESCOPES SPATIAL FILTER

We are currently studying a possible implementation of a two telescopes spatial filter (2TSF), with three additional requirements.
- The 2TSF must be achromatic. We want the possibility to use the largest possible spectral band permitted by the source, the detector and the atmosphere. For example a FT system using all or most of the flux in the J, H and K band, where we have low noise detectors, good adaptive optics and still relatively slow turbulence, would be particularly interesting to support the observations of MATISSE in the L, M and N bands.
- We want to minimize the number of measures or pixels necessary for the piston measurement, as this sets the ultimate sensitivity limit of the 2T FT module.
- We would like to have an unambiguous piston measure in a range larger than the fringe. The ideal solution would be a measure in a range of about 10 μm, approaching the overall atmospheric piston excursion at relatively short time scales.
- A secondary objective is to use both polarizations of light without needing polarization correctors, which have been used with success at the VLTI with PIONIER but might be difficult to implement for very broad bands.

All these requirements suggested us to investigate a bulk optics very broadband solution based on dispersed fringes, which is described in figure 3. A pair on input pupils produces a set of dispersed Fizeau fringes in the focal plane of a lens. Here we have simulated very broadband fringes from 1.05 to 2.45 μm. As the piston changes, the fringes drift in all spectral channels at different speeds and the dispersed fringe figure "bents". In the focal plane, we put an intensity mask, built from a "photograph" of the dispersed fringes at piston=0. The bright fringes coincide with a transmission part and the dark fringes with a reflective part. The transmitted flux C goes through a sharp maximum when the piston is zero and displays strong minima around ±1μm, as illustrated by figure 4a. The transmitted flux is anti-dispersed and collimated again. It is a fairly single-mode beam, with an average piston equal to the average of input pistons, as illustrated by figure 5 (which shows preliminary and not fully verified results). The rms of the output achromatic piston, over all wavelengths and other the double pupil is always smaller than 0.05 μm, i.e. λ/20 for the shortest wavelength.

When the input piston is non-zero, the first mask reflects bright fringes. A second mask can be used to analyze this reflected pattern. Here we have built the secondary masks in two parts: one half mimics the reflected pattern when the piston is 1 μm (the first maxima of reflected light), with transmission on the bright fringes and reflection on the dark fringes (at piston=1μm), while the second half is produced from the pattern reflected by the first mask when the input piston is -1μm, but the transmission is on the dark fringes and the reflection on the bright ones. This secondary mask results from an effort to optimize and symmetrize the piston estimator, with an unambiguous piston measure with the largest possible piston range. The secondary mask then transmits a beam in A and reflects one in B.

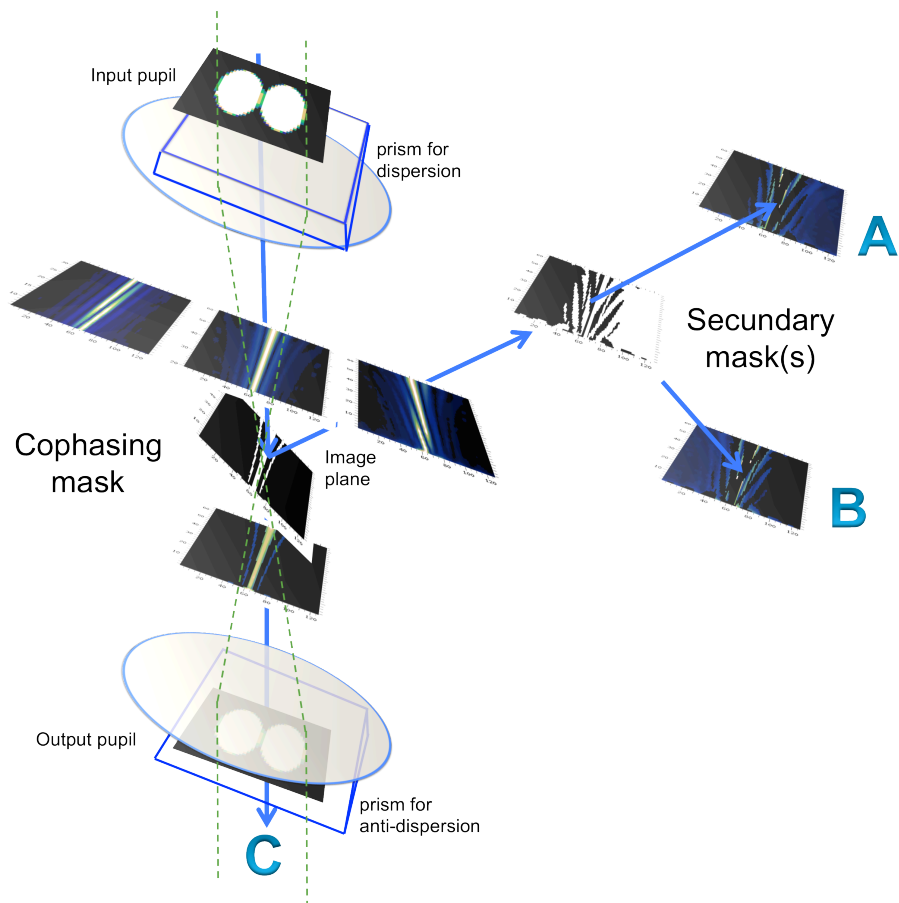

Figure 3: A broadband dispersed fringes achromatic 2 telescopes spatial filter and piston sensor.

In figure 4, we have plotted the total fluxes in beams A, B and C and the combination (A-B)/C that permits a piston estimate. The transmission of the 2TSF is of the order of 76% when the transmission and reflection surfaces of the grid are equivalent. This is comparable to the coupling efficiency of a single mode optical fiber, but in a much broader spectral range. If we want to estimate the piston from (A-B)/C, there is a bijective relation between this estimator and the piston in a ±1μm range. It is interesting to be able to measure the piston from very broadband fringes with only 3 pixels, but the unambiguous range is insufficient, and an optimum fringe tracker might need producing A and B in three well selected spectral bands, in an approach similar to the phase diversity methods[10]. Figure 4b also shows that the relation between (A-B)/C and the piston depends of the flux ratio between the two input beams. Although it might be possible to devise a fringe tracking strategy from an imprecise piston estimator (the slope of (A-B)/C change only by 20% when the flux ratio changes by 50%), A, B, and C do not allow to reconstruct the flux in beams 1 and 2, which is a condition for the hierarchical fringe tracking. This it is necessary to collect photometric beams at each one of the initial apertures. If all flux ratios are known, A, B, and C can be used for unambiguous piston estimates in a much broader range.

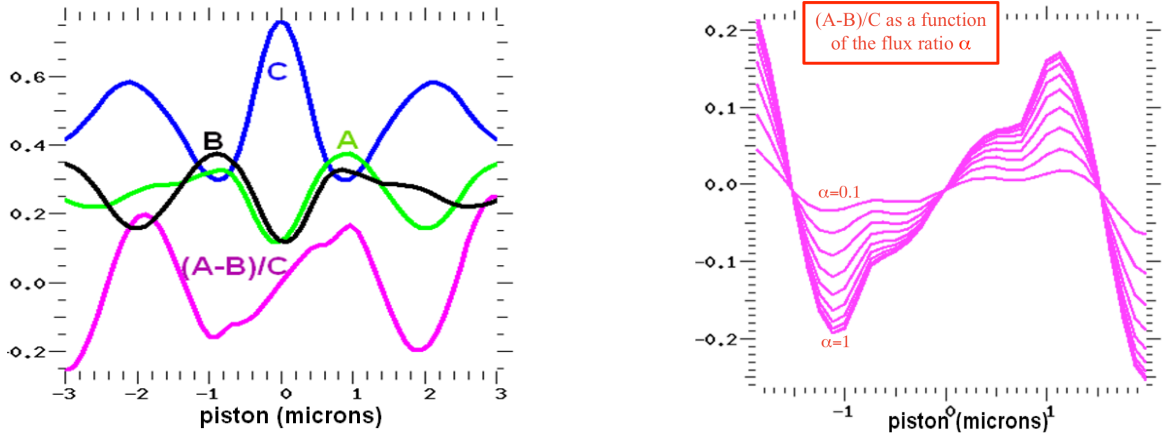

Figure 4: The intensity of beams produced by the dispersed fringes spatial filter and fringe sensor. In figure 4a (left), C represents the flux transmitted by the spatial filter that peaks at 76% when the input beams are cophased and stabilizes at 50% very far from cophasing. A and B are the intensity of the beams transmitted and reflected by the secondary grid. (A-B)/C is one of the possible measures that can be used to estimate the piston. The range for unambiguous piston estimate is of ± 1μm. Figure 4b (right) shows the variation of (A-B)/C as a function of the flux ratio α (α=1; 0.9, 0.8, 0.7, 0.6...).

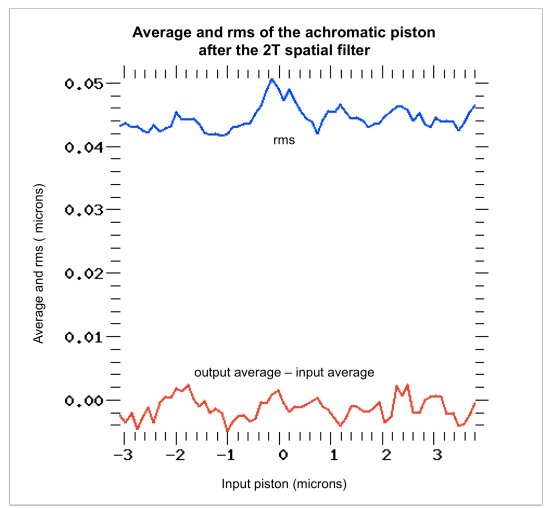

Figure 5: A first estimate of the spatial filter quality of the dispersed fringes spatial filter. The blue curve shows the piston rms, as a function of the input piston. It peaks at λ/20 for the shortest wavelength. The red curve shows the difference between the average output and input pistons.

WARNING: the results in this figure have been obtained very shortly before this manuscript closure date, and this feature of the simulation program has not been fully tested. So this result should be taken just as an indication of one of the ways to evaluate the spatial filtering quality, and as a first encouraging result.

## 5. CONCLUSION

We still have to optimize the parameters of the hierarchic fringe tracking approach and of all the 2TSF. We also have to simulate the control loop and to see how errors and loop breaks propagate. Then, we will realize a laboratory prototype.

However, we believe that our preliminary results give good indications that:
- Hierarchic fringe tracking is possible. Its sensitivity is similar to this of a two apertures fringe tracker, and does not decrease with the number of apertures. It might actually increase if we use the HFT structure to put the largest number of photons on the longest baselines with the lowest contrast.
- The basic element of an Hierarchic fringe tracker can be a spatial filter for pairs of telescopes, which produces a single mode beam from a pair of cophased telescope, with a coupling efficiency similar to this of a single aperture spatial filter.
- We have made a preliminary study on an achromatic very broadband two telescopes spatial filter. Its characteristics are promising but still have to be fully analyzed.
- It is possible to estimate the piston from very broadband fringes with only 3 to 5 pixels analyzing all the flux of two contributing beams, resulting from single apertures or from a group of apertures.
- The magnitude gain permitted by such a FT still has to be estimated exactly, when the parameters of the Hierarchic FT and of the 2TSF are optimized. However, from simple counts of the fluxes, the number of apertures and the number of measures, equation (4) gives a gain of 2.5 to 3.5 magnitudes with regard to the GRAVITY internal fringe tracker (assuming that the GRAVITY FT and the HFT differ only in the bandpass, beams and pixel management and that all other exposure time and control loop parameters are the same). If this is confirmed, it would place fringe tracking with the VLTI in the 13<K<14 range, with major applications on the AGN BLR program and many other science programs of GRAVITY and MATISSE.
- If the fringe tracker is used for off-axis cophasing, this gain of magnitude improves dramatically the sky coverage[11, 12]. For the VLTI, near the galactic pole it goes from less than 0.5% for K=10.5 to more than 7% for K=12.5. Within 20° of the galactic equator, the sky coverage would jump from about 15% to more than 100%.